\documentclass[reqno,11pt]{amsart}
\usepackage[utf8]{inputenc}
\usepackage{enumitem}
\usepackage{graphicx}
\usepackage{amscd}
\usepackage{slashed}
\usepackage{amssymb}
\usepackage{mathtools} 
\usepackage{pstricks}
\usepackage[mathscr]{eucal}
\numberwithin{equation}{section}
\allowdisplaybreaks[1]

\title[Two-Dimensional Area and Matter Flux in Causal Fermion Systems]{Two-Dimensional Area and Matter Flux in the Theory
of Causal Fermion Systems}

\author[E.\ Curiel]{Erik Curiel}
\address{Munich Center for Mathematical Philosophy \\ Ludwig-Maximilians-Universität \\ 80539 München \\ Germany}%
\email{erik@strangebeautiful.com}

\author[F.\ Finster]{Felix Finster}
\address{Fakult\"at f\"ur Mathematik \\ Universit\"at Regensburg \\ D-93040 Regensburg \\ Germany}%
\email{finster@ur.de}

\author[J.M.\ Isidro]{Jos{\'e} M. Isidro \\ \\ October 2019 / July 2020}
\address{Instituto Universitario de Matem\'atica Pura y Aplicada \\ Universidad Polit\'ecnica de Valencia \\
Valencia 46022 \\ Spain}
\email{joissan@mat.upv.es}

\newtheorem{Def}{Definition}[section]

\newtheorem{Prp}[Def]{Proposition}
\newtheorem{Lemma}[Def]{Lemma}

\newcommand{\Thanks}{\vspace*{.5em} \noindent \thanks}
\newcommand{\beq}{\begin{equation}}
\newcommand{\eeq}{\end{equation}}
\newcommand{\Proof}{\begin{proof}}
\newcommand{\QED}{\end{proof} \noindent}

\newcommand{\la}{\langle}
\newcommand{\ra}{\rangle}

\newcommand{\C}{\mathbb{C}}
\newcommand{\R}{\mathbb{R}}
\newcommand{\1}{\mbox{\rm 1 \hspace{-1.05 em} 1}}

\newcommand{\N}{\mathbb{N}}

\renewcommand{\H}{\mathscr{H}}

\newcommand{\bep}{\begin{pmatrix}}
\newcommand{\enp}{\end{pmatrix}}

\renewcommand{\O}{\mathscr{O}}

\newcommand{\F}{{\mathscr{F}}}

\renewcommand{\O}{{\mathscr{O}}}
\renewcommand{\L}{{\mathcal{L}}}
\newcommand{\Sact}{{\mathcal{S}}}
\newcommand{\s}{{\mathfrak{s}}}

\newcommand{\Lin}{\text{\rm{L}}}

\newcommand{\J}{\mathfrak{J}}
\newcommand{\Jin}{\mathfrak{J}^\text{\rm{\tiny{in}}}}
\newcommand{\Jlin}{\mathfrak{J}^\text{\rm{\tiny{lin}}}}
\newcommand{\Jtest}{\mathfrak{J}^\text{\rm{\tiny{test}}}}

\renewcommand{\div}{{\rm{div}}\,}

\DeclareFontFamily{OT1}{rsfso}{}
\DeclareFontShape{OT1}{rsfso}{m}{n}{ <-7> rsfso5 <7-10> rsfso7 <10-> rsfso10}{}
\DeclareMathAlphabet{\myscr}{OT1}{rsfso}{m}{n}

\newcommand\Felix[1]{}

\DeclareMathOperator{\Tr}{Tr}
\DeclareMathOperator{\tr}{tr}

\DeclareMathOperator{\supp}{supp}

\renewcommand{\u}{\mathfrak{u}}
\renewcommand{\v}{\mathfrak{v}}

\newcommand{\bitem}{\begin{itemize}[leftmargin=2.5em]}
\newcommand{\eitem}{\end{itemize}}

\begin{document}

\maketitle

\begin{abstract}
The notions of two-dimensional area, Killing fields and matter flux are introduced in the setting of
causal fermion systems. It is shown that for critical points of the causal action, the area
change of two-dimensional surfaces under a Killing flow in null directions is proportional
to the matter flux through these surfaces. This relation generalizes an equation in classical
general relativity due to Ted Jacobson to the setting of causal fermion systems.
\end{abstract}

\tableofcontents

\section{Introduction}
In 1995, Ted Jacobson derived the Einstein field equations from thermodynamic principles~\cite{jacobsonarea}.
At the heart of his argument is the formula
\beq \label{am}
\frac{d}{d\tau} A(S_\tau) = c\, F(S_\tau)
\eeq
which states that the area change of a family of two-surfaces~$S_\tau$ propagating along a null Killing direction
is proportional to the matter flux~$F(S_\tau)$ across these surfaces
(with~$c$ a constant). The relation~\eqref{am} has similarity to the Einstein equations
\beq \label{einstein}
R_{ij} - \frac{1}{2}\: R\: g_{ij} + \Lambda\: g_{ij} = 8 \pi G\: T_{ij}
\eeq
in that it gives a connection between geometry and matter. In order to make
this connection precise, Jacobson combined~\eqref{am} with the Raychaudhuri equation
to conclude that~\eqref{am} implies~\eqref{einstein}, up to an unspecified value of the cosmological
constant. Jacobson's starting point for the derivation of~\eqref{am} is the thermodynamic
formula~$\delta Q = T\, dS$ relating the heat flux to the change of entropy.
Using that the entropy of a horizon in thermal equilibrium is given by its area, the change of entropy
can be associated to an area change. Moreover, identifying the heat flux with the matter flux,
one obtains~\eqref{am}, with the proportionality constant~$c$ given by the
Hawking temperature of the horizon. 
The connection between thermodynamic concepts and the Einstein equations has been explored and discussed
extensively; see the review~\cite{padmanabhanreview} and references therein.

The theory of {\em{causal fermion systems}} is a recent approach to fundamental physics
where space-time is no longer modelled by a Lorentzian manifold but may instead have a
nontrivial, possibly discrete structure on a microscopic length scale which can be thought of
as the Planck scale (see the basics in Section~\ref{secprelim}, the reviews~\cite{dice2014, nrstg, review}, the textbook~\cite{cfs} or the website~\cite{cfsweblink}).
In the setting of causal fermion systems, the physical equations are formulated
via a variational principle, the {\em{causal action principle}}. The corresponding
Euler-Lagrange (EL) equations read (for details see the preliminaries in Section~\ref{secprelim})
\beq \label{EL}
\ell_\kappa |_M \equiv \inf_\F \ell_\kappa = 0 \:.
\eeq
In~\cite[Chapter~4]{cfs} it is shown that in a specific limiting case, the so-called
{\em{continuum limit}}, the EL equations give rise to the Einstein equations,
up to possible higher order corrections in curvature (which scale in powers of~$(\delta^2\:
\text{Riem})$, where~$\delta$ is the Planck length and~$\text{Riem}$ is the
curvature tensor). In this limiting case, space-time goes over to a Lorentzian manifold,
whereas the gravitational coupling constant~$G \sim \delta^2$ is determined by
the length scale~$\delta$ of the microscopic space-time structure.

The derivation of the Einstein equations in~\cite[Chapter~4]{cfs} has two disadvantages.
First, it is rather technical, because it relies on the detailed form of the regularized light-cone expansion of the
kernel of the fermionic projector. Consequently, the derivation does not give a good intuitive
understanding of the underlying mechanisms.
Second and more importantly, the Einstein equations are recovered
only in the continuum limit, but the methods do not give any insight into the geometric meaning
of the EL equations~\eqref{EL} for more general ``quantum'' space-times.
This raises the following question:
\begin{quote}
Given a general causal fermion system, how do the EL equations~\eqref{EL} relate matter to the
geometry of space-time?
\end{quote}
For a general causal fermion system we cannot work with tensor fields, making it impossible
to formulate the Einstein equations. There are general notions of connection and curvature~\cite{lqg},
but these geometric objects enter the EL equations in such an implicit way that so far it has not been possible
to give the EL equations a direct geometric interpretation.

Despite these principle difficulties, we here partially answer the above question by deriving an analog
of Jacobson's relation~\eqref{am} from the EL equations~\eqref{EL}. We thus establish a connection
between geometric properties (the change of the area of two-surfaces) and matter (the matter flux through
these surfaces).
In order to derive the Einstein equations, the remaining task would be to generalize the notion of null geodesics as well
as the Raychaudhuri equation to the setting of causal fermion systems. At present, this construction
can be carried out only in the continuum limit.
For the sake of greater generality and conceptual clarity, we here restrict attention to the derivation of~\eqref{am}.
In this way, for the first time we obtain a direct and intuitive understanding of how the EL equations~\eqref{EL}
relate matter fields to geometric quantities.
We point out that, in contrast to Jacobson's derivation, we do not use concepts of thermodynamics.
Indeed, at present it is unknown how ``entropy'' or ``temperature'' could be defined in the general setting of causal
fermion systems. But, turning Jacobson's arguments around, we do get a surprising connection between
the EL equations of the causal action principle and concepts from thermodynamics.
Exploring these connections more thoroughly and in more detail seems a promising objective for future research.

Our main task is to give the notions of ``two-dimensional area'' and ``matter flux'' a precise mathematical
meaning in the setting of causal fermion systems. Once this has been achieved,
the relation~\eqref{am} follows from the EL equations~\eqref{EL} by direct computation.
The starting point are surface layer integrals as introduced in~\cite{noether} which generalize the notion of
a three-dimensional surface integral to causal fermion systems.
Moreover, the conservation laws for such surface layer integrals derived in~\cite{jet, osi}
give analogs of the Green's formula and the Gau{\ss} divergence theorem.
In order to get from three-dimensional to two-dimensional surface integrals, for technical
simplicity we restrict attention to smooth space-times (see Definition~\ref{defsmooth})
and use so-called {\em{inner solutions}} (see Definition~\ref{definner}) in order to localize the
integration domain to the transverse intersection of a three-dimensional hypersurface and a surface layer
(see Figure~\ref{figarea} on page~\pageref{figarea}).

The paper is organized as follows. In Section~\ref{secprelim} we provide the necessary background
on causal fermion systems. Section~\ref{secinner} is devoted to defining inner solutions
and to collecting some of their properties. This makes it possible to define two-dimensional
{\em{area}} and {\em{area change}} (Section~\ref{secarea}). In Section~\ref{secflux} we introduce a notion of
{\em{Killing symmetries}} (see Definition~\ref{defkilling}) which is then used to define the {\em{matter flux}} through
a two-surface. In Section~\ref{seclightlike} we consider the limiting case that the Killing direction
becomes lightlike. In this limiting case, our formulas for the area change and the matter flux coincide,
giving~\eqref{am}.
A detailed appendix specifies the scalings of different contributions to the causal Lagrangian.
In particular, it is shown that the contributions by the matter fields are much smaller than the vacuum
contributions, a fact which is essential for our definition of Killing symmetries to be physically sensible.

\section{Preliminaries} \label{secprelim}

\subsection{Causal Fermion Systems and the Causal Action Principle}
\begin{Def} \label{defparticle} (causal fermion system) {\em{ 
Given a separable complex Hilbert space~$\H$ with scalar product~$\la .|. \ra_\H$
and a parameter~$n \in \N$ (the {\em{``spin dimension''}}), we let~$\F \subset \Lin(\H)$ be the set of all
self-adjoint operators on~$\H$ of finite rank, which (counting multiplicities) have
at most~$n$ positive and at most~$n$ negative eigenvalues. On~$\F$ we are given
a positive measure~$\rho$ (defined on a $\sigma$-algebra of subsets of~$\F$), the so-called
{\em{universal measure}}. We refer to~$(\H, \F, \rho)$ as a {\em{causal fermion system}}.
}}
\end{Def} \noindent
A causal fermion system describes a space-time together
with all structures and objects therein.
In order to single out the physically admissible
causal fermion systems, one must formulate physical equations. To this end, we impose that
the universal measure should be a minimizer of the causal action principle,
which we now introduce. For any~$x, y \in \F$, the product~$x y$ is an operator of rank at most~$2n$. 
However, in general it is no longer a selfadjoint operator because~$(xy)^* = yx$,
and this is different from~$xy$ unless~$x$ and~$y$ commute.
As a consequence, the eigenvalues of the operator~$xy$ are in general complex.
We denote these eigenvalues counting algebraic multiplicities
by~$\lambda^{xy}_1, \ldots, \lambda^{xy}_{2n} \in \C$
(more specifically,
denoting the rank of~$xy$ by~$k \leq 2n$, we choose~$\lambda^{xy}_1, \ldots, \lambda^{xy}_{k}$ as all
the non-zero eigenvalues and set~$\lambda^{xy}_{k+1}, \ldots, \lambda^{xy}_{2n}=0$).
We introduce the Lagrangian and the causal action by
\begin{align*}
\text{\em{Lagrangian:}} && \L(x,y) &= \frac{1}{4n} \sum_{i,j=1}^{2n} \Big( \big|\lambda^{xy}_i \big|
- \big|\lambda^{xy}_j \big| \Big)^2 \\ 
\text{\em{causal action:}} && \Sact(\rho) &= \iint_{\F \times \F} \L(x,y)\: d\rho(x)\, d\rho(y) \:. 
\end{align*}
The {\em{causal action principle}} is to minimize~$\Sact$ by varying the measure~$\rho$
under the following constraints:
\begin{align}
\text{\em{volume constraint:}} && \rho(\F) = \text{const} \quad\;\; & \label{volconstraint} \\
\text{\em{trace constraint:}} && \int_\F \tr(x)\: d\rho(x) = \text{const}& \label{trconstraint} \\
\text{\em{boundedness constraint:}} && \iint_{\F \times \F} 
|xy|^2
\: d\rho(x)\, d\rho(y) &\leq C \:, \label{Tdef}
\end{align}
where~$C$ is a given constant, $\tr$ denotes the trace of a linear operator on~$\H$, and
the absolute value of~$xy$ is the so-called spectral weight,
\[ |xy| := \sum_{j=1}^{2n} \big|\lambda^{xy}_j \big| \:. \]
This variational principle is mathematically well-posed if~$\H$ is finite-dimensional and if
one varies the measure in the class of regular Borel measures
(with respect to the topology on~$\Lin(\H)$ induced by the operator norm).
For the existence theory and the analysis of general properties of minimizing measures
we refer to~\cite{discrete, continuum, lagrange}.

Let~$\rho$ be a {\em{minimizing}} measure. {\em{Space-time}}
is defined as the support of this measure,
\[ 
M := \supp \rho \:, \]
Thus the space-time points are selfadjoint linear operators on~$\H$.
These operators contain a lot of additional information, which, if interpreted correctly,
gives rise to space-time structures like causal and metric structures, spinors
and interacting fields. This is explained in detail in~\cite[Chapter~1]{cfs}.

\subsection{The Euler-Lagrange Equations}
We now outline how the resulting Euler-Lagrange equations look like.
For technical simplicity, we work in the {\em{finite-dimensional}} setting, i.e.~$f:= \dim \H < \infty$.
Moreover, we assume that~$\rho$ is {\em{regular}} in the sense that all operators in~$M$
have exactly~$n$ positive and exactly~$n$ negative eigenvalues.
As is worked out in detail in~\cite{lagrange}, under certain technical assumptions
the constraints~\eqref{volconstraint}--\eqref{Tdef} can be treated as follows.
The trace constraint~\eqref{trconstraint} implies that the minimizing measure is
supported on operators of constant trace, i.e.\
\beq \label{trc}
\tr(x) = \text{c} \qquad \text{for all~$x \in M$}
\eeq
for some~$c>0$. The volume constraint~\eqref{volconstraint}
and the boundedness constraint~\eqref{Tdef}, on the other hand, can be taken into account
by positive Lagrange multipliers, which we denote by~$\s$ and~$\kappa$, respectively.
Thus we introduce the Lagrangian~$\L_\kappa$ by adding a Lagrange multiplier term to~
the causal Lagrangian,
\beq \label{Lkappa}
\L_\kappa \::\: \F \times \F \rightarrow \R\:,\qquad
\L_\kappa(x,y) := \L(x,y) + \kappa\: |xy|^2 \:.
\eeq
Moreover, we introduce the function~$\ell_\kappa$ by
\beq \label{ldef}
\ell_\kappa(x) = \int_M \L_\kappa(x,y)\: d\rho(y) - \s \:.
\eeq
Then the EL equations state that this function vanishes and is minimal on the support of~$\rho$,
\eqref{EL}.
For the derivation and technical details we refer to~\cite{jet}.

In view of the above regularity assumption as well as the fact that the trace is constant~\eqref{trc},
space-time~$M$ is a subset of the set~$\F^\text{reg}$ defined 
as the set of all operators on~$\H$ with the following properties:
\begin{itemize}[leftmargin=2em]
\item[(i)] $F$ is selfadjoint, has finite rank and (counting multiplicities) has
exactly~$n$ positive and~$n$ negative eigenvalues. \\[-0.8em]
\item[(ii)] The trace is constant, i.e.~$\tr(F) = c$.
\end{itemize}
The set~$\F^\text{reg}$ has a smooth manifold structure
(see the concept of a flag manifold in~\cite{helgason} or the detailed construction in~\cite[Section~3]{gaugefix}).
Therefore, working in~$\F^\text{reg}$, we are in the setting of causal variational principles
in the non-compact setting as introduced in~\cite[Section~2]{jet}.
Since in this paper we shall work with~$\F^\text{reg}$ throughout,
for notational simplicity we leave out the
superscript and denote the set of operators with the above properties~(i) and~(ii) again by~$\F$
(for more details on this procedure see also~\cite[Section~6.1]{perturb}).
For clarity, we point out that the smoothness of~$\F$ does not imply smoothness of~$M$.
Indeed, at this stage space-time~$M$ merely is a subset of~$\F$, but it does not need to be a
smooth submanifold of~$\F$ (smoothness of~$M$ will be introduced in Definition~\ref{defsmooth} below).

\subsection{The Weak Euler-Lagrange Equations and Jet Derivatives} \label{secwEL}
The EL equations~\eqref{EL} are nonlocal in the sense that
they make a statement on~$\ell_\kappa$ even for points~$x \in \F$ which
are far away from space-time~$M$.
It turns out that for the applications we have in mind, it is preferable to
evaluate the EL equations locally in a neighborhood of~$M$.
This concept leads to the {\em{weak EL equations}} introduced in~\cite[Section~4]{jet}.
We here give a slightly less general version of these equations which
is sufficient for our purposes. In order to explain how the weak EL equations come about,
we begin with the simplified situation that~$\ell_\kappa$ is a smooth function on~$\F$.
In this case, the minimality of~$\ell_\kappa$ implies that the derivative of~$\ell_\kappa$
vanishes on~$M$, i.e.\
\beq \label{ELweak}
\ell_\kappa|_M \equiv 0 \qquad \text{and} \qquad D \ell_\kappa|_M \equiv 0
\eeq
(where~$D \ell_\kappa(p) : T_p \F \rightarrow \R$ is the derivative).
In order to combine these two equations in a compact form,
it is convenient to consider a pair~$\u := (a, u)$
consisting of a real-valued function~$a$ on~$M$ and a vector field~$u$
on~$T\F$ along~$M$, and to denote the combination of 
multiplication and directional derivative by
\beq \label{Djet}
\nabla_{\u} \ell_\kappa(x) := a(x)\, \ell_\kappa(x) + \big(D_u \ell_\kappa \big)(x) \:.
\eeq
The equations~\eqref{ELweak} imply that~$\nabla_{\u} \ell_\kappa(x)$
vanishes for all~$\u$ and for all~$x \in M$.
The pair~$\u=(a,u)$ is referred to as a {\em{jet}}.

The Lagrangian~$\L_\kappa$~\eqref{Lkappa} of the causal action principle
is not everywhere smooth. Therefore, 
the directional derivative~$D_u \ell_\kappa$ in~\eqref{Djet} need not exist.
The method for dealing with this difficulty is to restrict attention to vector fields
for which the directional derivative is well-defined. We denote the resulting jet
space by
\[ \Jtest \;\subset\; \J := C^\infty(M, \R) \oplus \Gamma(M, T\F) \]
(where a function on~$M$ is by definition smooth if it is the restriction of a smooth function on~$\F$;
for technical details we refer to~\cite[Section~4]{jet} or~\cite[Section~2.2]{fockbosonic}).
Then the {\em{weak EL equations}} read
\beq \label{ELtest}
\nabla_{\u} \ell_\kappa|_M = 0 \qquad \text{for all~$\u \in \Jtest$}\:.
\eeq
For brevity, a solution of the weak EL equations is also referred to as a {\em{critical measure}}.

We finally note that, for the sake of technical simplicity, here we do not specify
the detailed regularity and smoothness assumptions but refer instead to~\cite[Section~2.2]{fockbosonic}.

\subsection{The Linearized Field Equations}
Usually, linearized fields are obtained by considering a family of
nonlinear solutions and linearizing with respect to a parameter~$\tau$
describing the field strength.
The analogous notion in the setting of causal fermion systems
is a linearization of a family of measures~$(\tilde{\rho}_\tau)$ which all satisfy the weak EL equations~\eqref{ELtest}.
It turns out to be fruitful to construct this family of measures by multiplying
a given critical measure~$\rho$ by a weight function~$f_\tau$ and then
``transporting'' the resulting measure with a mapping~$F_\tau$. More precisely, one considers the ansatz
\begin{align} \label{rhoFf}
\tilde{\rho}_\tau = (F_\tau)_* \big( f_\tau \, \rho \big) \:,
\end{align}
where~$f_\tau \in C^\infty(M, \R^+)$ and~$F_\tau \in C^\infty(M, \F)$ are smooth mappings,
and~$(F_\tau)_*\mu$ denotes the push-forward of a measure~$\mu$ (defined 
for a subset~$\Omega \subset \F$ by~$((F_\tau)_*\mu)(\Omega)
= \mu ( F_\tau^{-1} (\Omega))$; see for example~\cite[Section~3.6]{bogachev}).

The property of the family of measures~$\tilde{\rho}_\tau$ of the form~\eqref{rhoFf}
to satisfy the weak EL equation for all~$\tau$
means infinitesimally in~$\tau$ that the jet~$\v$ defined by
\[ 
\v = (b,v) := \frac{d}{d\tau} (f_\tau, F_\tau) \big|_{\tau=0} \]
satisfies the {\em{linearized field equations}} (for the derivation see~\cite[Section~3.3]{perturb}
or, in the simplified smooth setting, the textbook~\cite[Chapter~6]{intro})
\beq \label{eqlinlip}
\la \u, \Delta \v \ra|_M = 0 \qquad \text{for all~$\u \in \Jtest$} \:,
\eeq
where for any~$x \in M$,
\beq \label{Deldef}
\la \u, \Delta \v \ra(x) := \nabla_{\u} \bigg( \int_M \big( \nabla_{1, \v} + \nabla_{2, \v} \big) \L_\kappa(x,y)\: d\rho(y) - \nabla_\v \,\s \bigg)
\eeq
(and~$\nabla_1$ and~$\nabla_2$ act on the arguments~$x$ and~$y$ of the
Lagrangian, respectively).
We denote the vector space of all solutions of the linearized field equations by~$\Jlin \subset \J$.

\subsection{Surface Layer Integrals and Conservation Laws}
In the setting of causal fermion systems, the usual integrals over hypersurfaces in space-time are undefined.
Instead, one considers so-called {\em{surface layer integrals}}, being double integrals of the form
\beq \label{IntrOSI}
\int_\Omega d\rho(x) \int_{M \setminus \Omega} d\rho(y) \:(\cdots)\: \L_\kappa(x,y) \:,
\eeq
where~$\Omega$ is a subset of~$M$ and~$(\cdots)$ stands for a differential operator
acting on the Lagrangian. The structure of such surface layer integrals can be understood most easily 
in the special situation that the Lagrangian is of short range
in the sense that~$\L_\kappa(x,y)$ vanishes unless~$x$ and~$y$ are close together.
In this situation, we get a contribution to the double integral~\eqref{IntrOSI} only
if both~$x$ and~$y$ are close to the boundary~$\partial \Omega$.
With this in mind, surface layer integrals can be understood as an adaptation
of surface integrals to the setting of causal variational principles
(for a more detailed explanation see~\cite[Section~2.3]{noether}).

Surface layer integrals were first introduced in~\cite{noether} in order to
formulate Noether-like theorems for causal variational principles.
In particular, it was shown that there is a conserved
surface layer integral which generalizes the Dirac current
in relativistic quantum mechanics (see~\cite[Section~5]{noether}).
More recently, in~\cite{jet} another conserved
surface layer integral was discovered which gives rise to a symplectic form on the
solutions of the linearized field equations (see~\cite[Sections~3.3 and~4.3]{jet}).
A systematic study of conservation laws for surface layer integrals is given in~\cite{osi}.
The conservation law which is most relevant for our purposes is summarized in the next lemma.
\begin{Lemma} For any linearized solution~$\v \in \Jlin$,
\beq
\int_\Omega d\rho(x) \int_{M \setminus \Omega} d\rho(y)\: 
\big( \nabla_{1,\v} - \nabla_{2,\v} \big) \L_\kappa(x,y) 
= \int_\Omega \nabla_\v \, \s \: d\rho \:. \label{I1osi}
\eeq
\end{Lemma}
\Proof In view of the anti-symmetry of the integrand,
\[ \int_\Omega d\rho(x) \int_\Omega d\rho(y)\: 
\big( \nabla_{1,\v} - \nabla_{2,\v} \big) \L_\kappa(x,y) = 0 \:. \]
Adding this equation to the left side of~\eqref{I1osi}, we obtain
\begin{align*}
&\int_\Omega d\rho(x) \int_{M \setminus \Omega} d\rho(y)\: 
\big( \nabla_{1,\v} - \nabla_{2,\v} \big) \L_\kappa(x,y) \\
&= \int_\Omega d\rho(x) \int_M d\rho(y)\: 
\big( \nabla_{1,\v} - \nabla_{2,\v} \big) \L_\kappa(x,y) \\
&= \int_\Omega d\rho(x) \bigg( 2\, \nabla_\v \Big(\ell_\kappa(x) + \s \Big) 
- \big( \Delta \v \big)(x) - \nabla_\v \, \s \bigg) \:,
\end{align*}
where in the last line we used the definitions of~$\ell_\kappa$ and~$\Delta$
(see~\eqref{ldef} and~\eqref{Deldef}). Applying the weak EL equations~\eqref{ELtest}
and the linearized field equations~\eqref{eqlinlip} gives the result.
\QED

We shall also encounter surface layer integrals which depend on two jets~$\u$ and~$\v$
and have similarity to the surface layer integral
\beq \label{IOdef}
I_\Omega(\u, \v) := \int_\Omega d\rho(x) \int_{M \setminus \Omega} d\rho(y) \:\big(\nabla_{1,\u} - \nabla_{2,\u} \big)
\big(\nabla_{1,\v} + \nabla_{2,\v} \big) \L_\kappa(x,y) \:,
\eeq
which was introduced in~\cite{osi} and shown to satisfy a conservation law.
Symmetrizing and anti-symmetrizing in~$\u$ and~$\v$ gives the so-called {\em{surface layer inner product}}
and the {\em{symplectic form}}, respectively. 

\section{Inner Solutions} \label{secinner}
In preparation for introducing two-dimensional surface layer integrals,
we now construct a simple class of solutions of the linearized field equations.
These solutions do not have a dynamics of their own, but they
can be used for describing flows in space-time and for
``localizing'' objects on surface layers.

We again define {\em{space-time}}~$M:= \supp \rho \subset \F$ as the support of the universal measure.
Furthermore we make the following simplifying assumption:
\begin{Def} \label{defsmooth}
Space-time is {\bf{smooth and four-dimensional}} if~$M$ is a
four-dimensional smooth oriented submanifold of~$\F$. Moreover, in 
every local chart~$(x,U)$ of~$M$, the universal measure should be of the form
\beq \label{rhoh}
d\rho = h(x)\: d^4x \qquad \text{with~$h \in C^\infty(U, \R^+)$}\:.
\eeq
\end{Def} \noindent
From now on, we always assume that~$\rho$ is smooth and four-dimensional.

Let~$v \in \Gamma(M, TM)$ be a vector field. Then, under the above assumptions,
its {\em{divergence}} $\div v \in C^\infty(M, \R)$ can be defined by the relation
\[ \int_M \div v\: \eta(x)\: d\rho = -\int_M D_v \eta(x)\: d\rho(x) \:, \]
to be satisfied by all test functions~$\eta \in C^\infty_0(M, \R)$.
In a local chart~$(x,U)$, the divergence is computed by
\beq \label{divdef}
\div v = \frac{1}{h}\: \partial_j \big( h\, v^j \big)
\eeq
(where following the Einstein summation convention we sum over~$j=0,1,2,3$).

\begin{Def} \label{definner}
An {\bf{inner solution}} is a jet~$\v \in \J$
of the form
\[ \v = (\div v, v) \qquad \text{with} \qquad v \in \Gamma(M, TM) \:. \]
The vector space of all inner solution is denoted by~$\Jin \subset \J$.
\end{Def}

The name ``inner {\em{solution}}'' is justified by the following lemma:
\begin{Lemma} Every inner solution~$\v \in \Jin$
is a solution of the linearized field equations, i.e.
\[ \la \u, \Delta \v \ra_M = 0 \qquad \text{for all~$\u \in \Jtest$}\:. \]
\end{Lemma}
\Proof Applying the Gauss divergence theorem, one finds that for every~$f \in C^1_0(M, \R)$,
\[ \int_M \nabla_\v f\: d\rho = 
\int_M \big( \div v\: f + D_v f \big)\: d\rho 
= \int_M \div \big( f v \big)\: d\rho = 0 \:. \]
Likewise, in the linearized field equations we may integrate by parts in~$y$,
\begin{align*}
\la \u, \Delta \v \ra_M &= \nabla_\u \bigg( \int_M \big(\nabla_{1,\v} + \nabla_{2,\v} \big) \L_\kappa(x,y)
- \nabla_\v \,\s \bigg) \\
&= \nabla_\u \bigg( \int_M \nabla_{1,\v} \L_\kappa(x,y)
- \nabla_\v \,\s \bigg) \\
&= \nabla_\u \nabla_\v \ell_\kappa(x) = \nabla_\v \big( \nabla_\u \ell_\kappa(x) \big) 
- \nabla_{D_v \u} \ell_\kappa(x) = 0 \:.
\end{align*}
In the last step we used that~$\nabla_{D_v \u} \ell_\kappa(x)$ vanishes by the EL equations.
Moreover, the function~$\nabla_\u \ell_\kappa$ vanishes identically
on~$M$ in view of the weak EL equations. Therefore, it is differentiable in the direction
of every vector field on~$M$, and this directional derivative is zero.
\QED
For the sake of technical simplicity, here again we do not specify
the detailed regularity and smoothness assumptions but
refer instead to~\cite[Section~3]{fockbosonic}.

Inner solutions have the nice property that surface layer integrals simplify to
standard surface integrals, as is exemplified in the following lemma.
\begin{Def} Let~$\Omega \subset M$ be closed with smooth boundary~$\partial \Omega$.
On the boundary, we define the measure~$d\mu(\v,x)$ as the contraction of the volume form on~$M$
with~$v$, i.e.\ in local charts
\[ d\mu(\v,x) = h\: \epsilon_{ijkl} \:v^i\: dx^j dx^k dx^l \:, \]
where~$\epsilon_{ijkl}$ is the totally anti-symmetric Levi-Civita symbol
(normalized by~$\epsilon_{0123}=1$).
\end{Def}

\begin{Lemma} \label{lemmaintpart}
For every inner solution~$\v \in \Jin$ and any compact~$\Omega \subset M$,
\[ \int_\Omega d\rho(x) \int_{M \setminus \Omega} d\rho(y) \:\big(\nabla_{1,\v} - \nabla_{2,\v} \big)
\L_\kappa(x,y) = \s \int_{\partial \Omega} d\mu(\v,x) \:. \]
\end{Lemma}
\Proof Integrating by parts with the help of the Gau{\ss} divergence theorem, we obtain
\begin{align*}
&\int_\Omega d\rho(x) \int_{M \setminus \Omega} d\rho(y) \:\big(\nabla_{1,\v} - \nabla_{2,\v} \big)
\L_\kappa(x,y) \\
&= \int_{\partial \Omega} d\mu(\v,x) \int_{M \setminus \Omega} d\rho(y)\:
\L_\kappa(x,y) + \int_\Omega d\rho(x) \int_{\partial \Omega} d\mu(\v,y)\: \L_\kappa(x,y) \\
&= \int_{\partial \Omega} d\mu(\v,x) \int_M d\rho(y)\: \L_\kappa(x,y)\:,
\end{align*}
where in the last step we used the symmetry of~$\L_\kappa$.
Employing the EL equations gives the result.
\QED

\section{Two-Dimensional Area and Area Change} \label{secarea}
\begin{figure}
%
\psscalebox{1.0 1.0} 
{
\begin{pspicture}(0,-1.6575384)(7.317311,1.6575384)
\definecolor{colour0}{rgb}{0.8,0.8,0.8}
\definecolor{colour2}{rgb}{0.6,0.6,0.6}
\definecolor{colour1}{rgb}{0.4,0.4,0.4}
\pspolygon[linecolor=colour0, linewidth=0.02, fillstyle=solid,fillcolor=colour0](1.9208822,-1.632477)(3.1051679,-1.637477)(3.1358821,-1.0824771)(3.2380252,-0.4174771)(3.1915965,0.1575229)(3.0908823,0.5975229)(3.0208821,1.0525229)(2.987311,1.627523)(1.7408823,1.6475229)(0.03445365,1.6475229)(0.046596505,0.4775229)(0.033739362,-0.3024771)(0.03231079,-0.8474771)(0.03588222,-1.3574771)(0.03231079,-1.637477)
\pspolygon[linecolor=colour2, linewidth=0.04, fillstyle=solid,fillcolor=colour2](0.03231079,-0.8424771)(0.6273108,-0.8424771)(0.8823108,-0.8174771)(1.3173108,-0.7674771)(1.7223108,-0.7274771)(2.4573107,-0.6824771)(3.0773108,-0.6774771)(3.8073108,-0.7274771)(4.552311,-0.8174771)(5.092311,-0.9124771)(5.6873107,-1.012477)(6.2273107,-1.0574771)(6.442311,-1.0674771)(6.442311,-1.637477)(0.05231079,-1.627477)
\psbezier[linecolor=colour1, linewidth=0.04](2.987311,1.6425229)(2.9650757,0.8289815)(3.3506305,-0.2659784)(3.2073107,-0.7875877355247934)(3.063991,-1.3091971)(3.0998464,-1.4602805)(3.0873108,-1.637477)
\rput[bl](0.15731078,1.2975229){\normalsize{$V$}}
\psbezier[linecolor=black, linewidth=0.04](0.01731079,-0.8274771)(1.0182934,-0.8497122)(1.8987651,-0.5991574)(3.339531,-0.6674770927429199)(4.780297,-0.7357968)(5.0838623,-0.9849415)(6.4523106,-1.0574771)
\rput[bl](5.9973106,-1.4874771){\normalsize{$\Omega$}}
\psbezier[linecolor=black, linewidth=0.02](0.002310791,-0.2624771)(1.1932933,-0.5447122)(1.7887651,0.24084264)(3.324531,-0.10247709274291993)(4.860297,-0.44579682)(5.0688624,-0.41994146)(6.4373107,-0.4924771)
\rput[bl](6.7273107,-0.6424771){\normalsize{$\partial \Omega_\tau$}}
\pscircle[linecolor=black, linewidth=0.04, fillstyle=solid,fillcolor=black, dimen=outer](3.217311,-0.6624771){0.095}
\pscircle[linecolor=black, linewidth=0.04, fillstyle=solid,fillcolor=black, dimen=outer](3.2023108,-0.08247709){0.095}
\rput[bl](3.342311,-0.5974771){\normalsize{$S$}}
\rput[bl](3.3173108,0.017522907){\normalsize{$S_\tau$}}
\psline[linecolor=black, linewidth=0.02, arrowsize=0.05291667cm 2.0,arrowlength=1.4,arrowinset=0.0]{->}(3.0223107,0.49752292)(2.9573107,0.9175229)
\psline[linecolor=black, linewidth=0.02, arrowsize=0.05291667cm 2.0,arrowlength=1.4,arrowinset=0.0]{->}(2.8823109,1.107523)(2.8373108,1.5275229)
\rput[bl](2.5023108,1.2975229){\normalsize{$v$}}
\psline[linecolor=black, linewidth=0.02, arrowsize=0.05291667cm 2.0,arrowlength=1.4,arrowinset=0.0]{->}(2.1423109,-0.6874771)(2.0373108,-0.2874771)
\psline[linecolor=black, linewidth=0.02, arrowsize=0.05291667cm 2.0,arrowlength=1.4,arrowinset=0.0]{->}(1.2773108,-0.7574771)(1.1973108,-0.4324771)
\psline[linecolor=black, linewidth=0.02, arrowsize=0.05291667cm 2.0,arrowlength=1.4,arrowinset=0.0]{->}(0.5273108,-0.8224771)(0.5223108,-0.47247708)
\psline[linecolor=black, linewidth=0.02, arrowsize=0.05291667cm 2.0,arrowlength=1.4,arrowinset=0.0]{->}(2.7873108,-0.032477092)(2.717311,0.37252292)
\psline[linecolor=black, linewidth=0.02, arrowsize=0.05291667cm 2.0,arrowlength=1.4,arrowinset=0.0]{->}(2.0123107,-0.07747709)(1.9273108,0.40252292)
\psline[linecolor=black, linewidth=0.02, arrowsize=0.05291667cm 2.0,arrowlength=1.4,arrowinset=0.0]{->}(2.8673108,-0.6374771)(2.852311,-0.1574771)
\psline[linecolor=black, linewidth=0.02, arrowsize=0.05291667cm 2.0,arrowlength=1.4,arrowinset=0.0]{->}(0.4473108,-0.3274771)(0.3973108,0.1675229)
\psline[linecolor=black, linewidth=0.02, arrowsize=0.05291667cm 2.0,arrowlength=1.4,arrowinset=0.0]{->}(1.1473107,-0.2774771)(1.0623108,0.2025229)
\psline[linecolor=black, linewidth=0.02, arrowsize=0.05291667cm 2.0,arrowlength=1.4,arrowinset=0.0]{->}(3.3023107,0.76252294)(3.237311,1.2175229)
\psline[linecolor=black, linewidth=0.02, arrowsize=0.05291667cm 2.0,arrowlength=1.4,arrowinset=0.0]{->}(3.6873107,-0.6774771)(3.6923108,-0.2574771)
\psline[linecolor=black, linewidth=0.02, arrowsize=0.05291667cm 2.0,arrowlength=1.4,arrowinset=0.0]{->}(3.987311,-0.2374771)(3.967311,0.18252291)
\psline[linecolor=black, linewidth=0.02, arrowsize=0.05291667cm 2.0,arrowlength=1.4,arrowinset=0.0]{->}(3.7823107,0.4125229)(3.737311,0.8325229)
\psline[linecolor=black, linewidth=0.02, arrowsize=0.05291667cm 2.0,arrowlength=1.4,arrowinset=0.0]{->}(2.9223108,-1.497477)(2.977311,-1.0924771)
\psline[linecolor=black, linewidth=0.02, arrowsize=0.05291667cm 2.0,arrowlength=1.4,arrowinset=0.0]{->}(3.2923107,-1.377477)(3.3473108,-0.9274771)
\end{pspicture}
}
\caption{Two-dimensional area and area change.}
\label{figarea}
\end{figure}
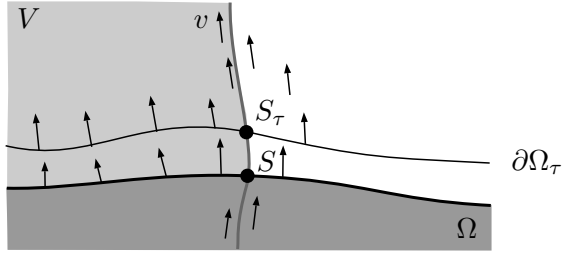

In what follows, we assume that~$\rho$ is a critical measure and that
the corresponding space-time~$M:=\supp \rho$ is smooth and four-dimensional (see Definition~\ref{defsmooth}).
Let~$V \subset M$ be an open subset of space-time with smooth boundary~$\partial V$.
Moreover, we let~$v$ be a vector field on~$M$ which is tangential to~$\partial V$
and denote the corresponding inner solution by~$\v=(b:=\div v, v)$.
Moreover, we let~$\Omega \subset M$ be another open set. We assume that the boundaries of~$\Omega$
and~$V$ intersect transversely, meaning that
\[ S := \partial \Omega \cap \partial V \]
is a two-dimensional surface (see Figure~\ref{figarea}).
We define its area by
\[ A := \int_{\partial \Omega \cap V} d\mu(\v, x) \int_{M \setminus V} d\rho(y)\: \L_\kappa(x,y) \:. \]
Again integrating by parts and assuming that the vector field~$v$ vanishes in the past,
this area can be written alternatively as
\begin{align}
A &= \int_{\Omega \cap V} d\rho(x) \:\nabla_\v \int_{M \setminus V} d\rho(y)\: \L_\kappa(x,y) \notag \\
&= \int_{\Omega \cap V} d\rho(x) \int_{M \setminus V} d\rho(y)\: \big( \nabla_{1,\v} \pm \nabla_{2,\v} \big) \L_\kappa(x,y)
\label{Aform}
\end{align}
(the notation~$\pm$ means that the formula holds for either choice of the sign;
this is because the corresponding terms vanishes after integrating by parts in view of
Lemma~\ref{lemmaintpart} and the fact that~$v$ is tangential to~$\partial V$).
Before going on, we point out that our definition of area involves a scaling factor coming from
the Lagrangian (in Appendix~\ref{appA}, this scaling factor is computed and given in~\eqref{nuval}).

The vector field also defines a flow of surfaces. Indeed, let~$\Phi_\tau$ the diffeomorphism generated by the
vector field~$v$ and~$\Omega_\tau := \Phi_\tau(\Omega)$. Then the surface~$S$ flows to
\[ S_\tau := \Phi_\tau(A) = \partial \Omega_\tau \cap \partial V \:. \]
The area change is obtained by differentiating~\eqref{Aform}
\begin{Prp} \label{prpareachange}
The infinitesimal change of the area~\eqref{Aform} in the direction of the
vector field~$v$ is given by
\begin{align}
&\frac{d}{d\tau} A(S_\tau) \Big|_{\tau=0} = \int_{\Omega \cap V} d\rho(x) \int_{M \setminus V} d\rho(y)\: 
\big( \nabla_{1,\v} + \nabla_{2,\v} \big) \big( \nabla_{1,\v} - \nabla_{2,\v} \big) \L_\kappa(x,y) \label{dtA1} \\
&\quad + \int_{\Omega \cap V} d\rho(x) \int_{M \setminus V} d\rho(y)\:  \L_\kappa(x,y)\:
\Big( D_v \div v(x) - D_v \div v(y) \Big) \label{dtA2} \\
&\quad + \int_{\Omega \cap V} d\rho(x) \int_{M \setminus V} d\rho(y) \:\big( \nabla_{1,\v} - \nabla_{2,\v} \big) \L_\kappa(x,y) \: \big( \div v(x) + \div v(y) \big) \:. \label{dtA3}
\end{align}
\end{Prp} 
\Proof
The inner solution can be written as~$\v = (\div v, \partial_\tau)$.
When differentiating~\eqref{Aform} with respect to~$\tau$, one must 
take into account that, according to~\eqref{rhoh} and~\eqref{divdef}, the measure~$d\rho$
depends smoothly on~$\tau$ and that its $\tau$-derivative is the signed measure~$\div v\, d\rho$.
Moreover, one must differentiate the factors~$\div v$ in the scalar component of~$\v$.
This gives the result.
\QED

\section{Killing Fields and Matter Flux} \label{secflux}
In differential geometry, a Killing field describes a symmetry of the metric tensor.
Here instead of the metric we must work with the structures of the causal fermion system:
the measure~$\rho$ and the Lagrangian~$\L$. In view of~\eqref{rhoh} and~\eqref{divdef},
the condition that~$\rho$ should be invariant in the direction of the vector field~$v$
simply means that~$\div v=0$. Then the corresponding inner solution has a vanishing scalar component,
\[ \v = (0, v) \:. \]
A symmetry of the Lagrangian, on the other hand, is
captured in the expression~$(D_{1,v} + D_{2,v})\L(x,y)$ where both arguments are differentiated
in the direction of the vector field~$v$. Since the Lagrangian also involves the matter fields,
it would be a too strong condition to demand that this expression is zero. Instead, this expression
must be sufficiently small, as stated in the next definition and worked out in detail in
the appendix (where~$m$ is the mass of the Dirac particles of the system,
$\delta$ is the Planck length, $\varepsilon$ is the regularization
length, and for convenience we chose the scaling parameters~$\sigma=\lambda=1$).
\begin{Def} \label{defkilling}
A vector field~$v$ on~$M$ is called {\bf{Killing field}} of the causal fermion system if
the following conditions hold:
\bitem
\item[(i)] The divergence~\eqref{divdef} of~$v$ vanishes,
\[ \div v = 0 \:. \]
\item[(ii)] The directional derivative of the Lagrangian is small in the sense that
\beq \label{DvL}
\big( D_{1,v} + D_{2,v} \big) \L_\kappa(x,y) \lesssim \frac{m^4}{\varepsilon^4\: \delta^4} \:.
\eeq
\eitem
\end{Def}

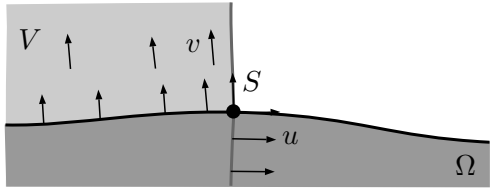
\begin{figure}
%
\psscalebox{1.0 1.0} 
{
\begin{pspicture}(-3.5,-1.2303988)(10.170529,1.2303988)
\definecolor{colour0}{rgb}{0.8,0.8,0.8}
\definecolor{colour1}{rgb}{0.6,0.6,0.6}
\definecolor{colour2}{rgb}{0.4,0.4,0.4}
\pspolygon[linecolor=colour0, linewidth=0.02, fillstyle=solid,fillcolor=colour0](1.9141006,-1.2053374)(3.0983863,-1.2103374)(3.0241005,-0.6453374)(3.0312436,0.0)(3.0148149,0.5596626)(3.0091007,0.7396626)(3.0041006,0.9946626)(3.0055292,1.2196625)(1.5091006,1.2096626)(0.032672033,1.1946626)(0.03981489,0.9046626)(0.026957747,0.12466259)(0.025529174,-0.4203374)(0.029100602,-0.9303374)(0.025529174,-1.2103374)
\pspolygon[linecolor=colour1, linewidth=0.04, fillstyle=solid,fillcolor=colour1](0.020529175,-0.4153374)(0.6155292,-0.4153374)(0.8705292,-0.3903374)(1.3055291,-0.3403374)(1.7105292,-0.3003374)(2.4455292,-0.25533742)(3.065529,-0.25033742)(3.7955291,-0.3003374)(4.5405293,-0.3903374)(5.080529,-0.4853374)(5.675529,-0.5853374)(6.215529,-0.6303374)(6.430529,-0.6403374)(6.430529,-1.2103374)(0.040529177,-1.2003374)
\psbezier[linecolor=colour2, linewidth=0.04](3.0105293,1.2296625)(2.994465,0.7911212)(3.0596163,-0.02383872)(3.0405293,-0.360448052755018)(3.021442,-0.69705737)(3.0112207,-1.0331409)(3.007802,-1.2103374)
\rput[bl](0.18552917,0.6046626){\normalsize{$V$}}
\psbezier[linecolor=black, linewidth=0.04](0.0105291745,-0.4003374)(1.0115118,-0.4225725)(1.8919835,-0.17201768)(3.3327494,-0.24033740997314454)(4.773515,-0.30865714)(5.0770807,-0.5578018)(6.445529,-0.6303374)
\rput[bl](5.990529,-1.0603374){\normalsize{$\Omega$}}
\pscircle[linecolor=black, linewidth=0.04, fillstyle=solid,fillcolor=black, dimen=outer](3.0405293,-0.2203374){0.095}
\rput[bl](3.1555293,0.04966259){\normalsize{$S$}}
\psline[linecolor=black, linewidth=0.02, arrowsize=0.05291667cm 2.0,arrowlength=1.4,arrowinset=0.0]{->}(2.6955292,-0.22533742)(2.6555293,0.21966259)
\rput[bl](2.4105291,0.5646626){\normalsize{$v$}}
\psline[linecolor=black, linewidth=0.02, arrowsize=0.05291667cm 2.0,arrowlength=1.4,arrowinset=0.0]{->}(2.1355293,-0.2603374)(2.105529,0.14466259)
\psline[linecolor=black, linewidth=0.02, arrowsize=0.05291667cm 2.0,arrowlength=1.4,arrowinset=0.0]{->}(1.2705292,-0.3303374)(1.2455292,0.06966259)
\psline[linecolor=black, linewidth=0.02, arrowsize=0.05291667cm 2.0,arrowlength=1.4,arrowinset=0.0]{->}(0.52052915,-0.3953374)(0.49552917,0.00966259)
\psline[linecolor=black, linewidth=0.02, arrowsize=0.05291667cm 2.0,arrowlength=1.4,arrowinset=0.0]{->}(2.7805293,0.3946626)(2.740529,0.8746626)
\psline[linecolor=black, linewidth=0.02, arrowsize=0.05291667cm 2.0,arrowlength=1.4,arrowinset=0.0]{->}(2.0055292,0.3496626)(1.9555292,0.7746626)
\psline[linecolor=black, linewidth=0.02, arrowsize=0.05291667cm 2.0,arrowlength=1.4,arrowinset=0.0]{->}(3.0455291,-0.18533741)(3.0305293,0.2946626)
\psline[linecolor=black, linewidth=0.02, arrowsize=0.05291667cm 2.0,arrowlength=1.4,arrowinset=0.0]{->}(3.0053284,-1.017598)(3.59073,-1.0230768)
\psline[linecolor=black, linewidth=0.02, arrowsize=0.05291667cm 2.0,arrowlength=1.4,arrowinset=0.0]{->}(0.87552917,0.34466258)(0.83552915,0.8146626)
\psline[linecolor=black, linewidth=0.02, arrowsize=0.05291667cm 2.0,arrowlength=1.4,arrowinset=0.0]{->}(3.0253847,-0.58398014)(3.6106737,-0.59669465)
\psline[linecolor=black, linewidth=0.02, arrowsize=0.05291667cm 2.0,arrowlength=1.4,arrowinset=0.0]{->}(3.0804245,-0.2433218)(3.665634,-0.227353)
\rput[bl](3.690529,-0.6503374){\normalsize{$u$}}
\end{pspicture}
}
\caption{Matter flux through~$S$.}
\label{figflux}
\end{figure}

\Felix{Maybe we should point out that in~\eqref{DvL} we make use of the fact
that we have a {\em{critical}} measure. Indeed, if the EL equations
corresponding to the causal action principle were violated, in~\eqref{DvL} we would get a much bigger contribution!}
The contributions allowed in~\eqref{DvL} describe the
derivative of the matter content in the direction of~$v$
(for more details and computations in the static case see~\cite{pmt}).
Next, we let~$u$ be a vector field which is tangential to~$\partial \Omega$
(see Figure~\ref{figflux}) and~$\u=(\div u, u)$ the corresponding inner solution.
Then the matter flux through the surface~$S$ in direction~$u$ can be defined 
in analogy to~\eqref{IOdef} by
\beq \label{FOdef}
F(S_\tau) := \int_{\Omega \cap V} d\rho(x) \int_{M \setminus V} d\rho(y) \:\big(\nabla_{1,\u} - \nabla_{2,\u} \big)
\big(\nabla_{1,\v} + \nabla_{2,\v} \big) \L_\kappa(x,y) \:.
\eeq

\section{The Limiting Case of Lightlike Propagation} \label{seclightlike}
We now consider the limiting case of a lightlike Killing direction.
By this we simply mean the limiting case that the vector fields~$v$ and~$u$ coincide.
Then the divergence and its derivatives in~\eqref{dtA2} and~\eqref{dtA3} vanish.
As a consequence, we also obtain agreement between the formulas for the change of area
in Proposition~\ref{prpareachange} and the matter flux in~\eqref{FOdef},
\beq \label{result}
\frac{d}{d\tau} A(S_\tau) = F(S_\tau) \:.
\eeq
This generalizes the analogous formula used by Ted Jacobson~\eqref{am}.
We note for clarity that the constant~$c$ in~\eqref{am} is explicitly contained
in~\eqref{result} as well. Indeed, the
orders in~$\varepsilon$ and~$\delta$ in~\eqref{DvL} together with the scaling of the causal
Lagrangian determines this constant. We do not enter the details because it is
clear already from dimensional consideration and from consistency to the
Einstein equations as derived in~\cite[Chapter~4]{cfs} that the resulting
gravitational constant scales like~$G \sim \delta^2$.

\appendix
\section{Scaling of the Causal Lagrangian} \label{appA}
The goal of this appendix is to derive the scaling of the contributions on the right side
of the Killing equation~\eqref{DvL}. To this end, we analyze the scaling of the causal Lagrangian
and of the constraints both in the Minkowski vacuum and in the presence of matter.
Our limited knowledge on the microscopic structure of space-time will be reflected in
a number of unknown parameters (denoted by~$p$, $q$ and~$\hat{q}$).
As we shall see, the scalings in~\eqref{DvL}
are universal in the sense that they are independent of these unknowns.

\subsection{Relevant Length Scales}
We recall the length scales which enter the construction of causal fermion systems.
We always work in natural units where~$\hbar=c=1$. Then the gravitational coupling
constant~$G$ has dimension length squared. The corresponding length scale is the
\[ \text{\em{Planck length}} \qquad \delta \approx 1.6\cdot 10^{-35}\,\text{meters} \:. \]
The rest mass of the Dirac particles determines another length scale,
the {\em{Compton length}}~$m^{-1}$. Next, there is the {\em{regularization length}}~$\varepsilon$.
The simplest and most natural assumption is to identify the regularization length with the Planck length.
However, as is explained in detail in~\cite[Chapter~4]{cfs}, this assumption is too naive,
because the regularization length should be
much smaller than the Planck length. Therefore, we must treat~$\varepsilon$
and~$\delta$ as different parameters. We merely assume that
\[ \varepsilon \ll \delta \ll \frac{1}{m}\:. \]
Finally, there is the length scale~$l_{\text{\tiny{macro}}}$ of macroscopic physics.
Clearly, this length scale depends on the physical system under considerations. Since
energies much larger than the rest masses of the heaviest fermions are not accessible to
experiments, we always assume that
\[ \frac{1}{m} \lesssim l_{\text{\tiny{macro}}}\:. \]

\subsection{Freedom in Rescaling Solutions of the Euler-Lagrange Equations} \label{secrescale}
Let~$\rho$ be a critical measure of the causal action principle. Then for suitable Lagrange multipliers~$c, \s>0$,
the equations~\eqref{trc} and~\eqref{EL} hold. We now write the function~$\ell_\kappa$
defined by~\eqref{ldef} as
\[ 
\ell_\kappa(x) := \ell(x) + \kappa \,\mathfrak{t}(x)\:, \]
where
\begin{align*}
\ell(x) &:= \int_M \L_\kappa(x,y)\: d\rho(y)  - \s \\
\mathfrak{t}(x) &:= \int_M |xy|^2\: d\rho(y) \:.
\end{align*}

There is a two-parameter family of rescalings which again give critical measures.
Indeed, the new measure~$\tilde{\rho}$ defined by
\beq \label{tilrho}
\tilde{\rho}(\Omega) = \sigma\: \rho\Big( \frac{\Omega}{\lambda} \Big) \qquad \text{with~$\lambda, \sigma>0$}\:,
\eeq
again satisfies the EL equation with new Lagrange multipliers
\[ \tilde{c} = \lambda\, c \quad \text{and} \qquad \tilde{\s} = \sigma\, \lambda^4\: \s\:. \]
This rescaling freedom could be fixed for example by imposing that
\[ c = \s = 1 \:. \]
Note that the Lagrange multiplier~$\kappa$ remains unchanged; it is a dimensionless
parameter which characterizes the solution independent of the values of~$c$ and~$\s$.
For what follows, it is preferable {\em{not}} to fix this rescaling freedom,
because this simplifies the comparison of our formulas with the computations
in~\cite{reg, action}.

\subsection{Scalings in the Minkowski Vacuum} \label{secmink}
For most of the following formulas, it makes no difference how we regularize.
Therefore, we mainly work in the $i \varepsilon$-regularization introduced in~\cite[Section~2.4.1]{cfs}.
Whenever the form of the regularization does matter, we shall discuss the results of the
paper~\cite{reg} where the regularization effects were analyzed for a general class of regularizations.
Implementing the scaling freedom in~\eqref{tilrho}, the kernel of the fermionic project takes the form
\[ P(x,y) \simeq \lambda \int \frac{d^4k}{(2 \pi)^4}\: (\slashed{k} + m) \,\delta(k^2-m^2)\, \Theta(-k^0)
\: \exp \big( \varepsilon k^0 \big)\: e^{-ik(x-y)} \:, \]
where~$\varepsilon>0$ is the regularization length. Power counting shows that this distribution
has length dimension
\[ 
P(x,y) \simeq \frac{\lambda}{l^3} \]
(where~$l$ has dimension length). We infer that the closed chain, its eigenvalues and
the Lagrangian scale like
\begin{align}
A_{xy},\: \lambda^{xy}_i &\sim \frac{\lambda^2}{l^6} \label{lamscale} \\
\L(x,y),\: \sum_{i=1}^{2n} \big|\lambda^{xy}_i \big|^2 &\sim \frac{\lambda^4}{l^{12}} \:. \label{Lscale}
\end{align}
In particular, the local trace scales like
(for details see~\cite[Section~2.5]{cfs})
\beq \label{trx}
\tr(x) = \Tr \big( P(x,x) \big) \simeq \frac{\lambda m}{\varepsilon^2} \:.
\eeq

Following the constructions in~\cite[Section~1.2]{cfs}, the measure~$\rho$ is chosen as the push-forward
of the Lebesgue measure on Minkowski space. Taking into account the scaling in~\eqref{tilrho}, we set
\[ \tilde{\rho} = F^\varepsilon_* (\sigma\, \mu) \qquad \text{with} \qquad d\mu=d^4x\:. \]
The Lagrange multiplier term involving~$\kappa$ is non-zero even in the massless case.
Therefore, 
its scaling can be determined by power counting,
\beq \label{kappacount}
\kappa\, \mathfrak{t}(x) \simeq \kappa\:
\frac{\sigma\, \lambda^4}{\varepsilon^8}\:.
\eeq
In more detail, this scaling behavior is obtained in the formalism of the continuum limit
(as introduced in~\cite[Sections~2.4, 3.5 and~4.2]{cfs}) by writing
\beq \label{Txyform}
\sum_{i=1}^{2n} \big|\lambda^{xy}_i \big|^2 \simeq \lambda^2\: (\deg = 6) \:,
\eeq
Here the factor~$(\deg = 6)$ is a contribution which is localized on the light cone and
has length dimension~$l^{-12}$. It can be written more explicitly as
\beq
(\deg=6) =
\frac{h(\xi)}{(\varepsilon t)^5}\: \delta_\varepsilon(\xi^2)\: \epsilon(\xi^0) \:, \label{znotation}
\eeq
where~$\xi^2 := \la \xi,\xi \ra$ denotes the Minkowski inner product,
$\delta_\varepsilon$ is a function which in the limit~$\varepsilon \searrow 0$
converges to the $\delta$-distribution, and~$\epsilon$ is the step function.
Moreover, the function~$h$ is smooth away from the origin and bounded
from above and below, uniformly in~$\varepsilon$.
Integrating in~\eqref{Txyform} and~\eqref{znotation} over~$y$ gives the scaling~\eqref{kappacount}.

The scaling of the Lagrangian is more subtle because there are different contributions involving different
powers in the mass. We discuss them after each other.

\subsubsection{Contributions Away from the Light Cone}
Away from the light cone (i.e.\ if~$(y-x)^2 \neq 0$), the fermionic projector is smooth,
so that the Lagrangian is well-defined without a regularization.
The resulting contribution to~$\ell$ is bounded.
Expanding in powers of the mass, the lowest order is given by~$\L(x,y) \sim m^6$
(for details see~\cite{vacstab} and~\cite{reg}).
If this contribution extended up to the light cone, the corresponding contribution to~$\ell$
could be computed similar to~\eqref{kappacount} with power counting to be
\[ \ell(x)+ \s = \int_M \L(x,y)\: d\rho(y) \simeq\frac{\sigma\, \lambda^4\, m^6}{\varepsilon^2}\:. \]
However, this picture is not quite correct because the detailed analysis in~\cite{reg}
reveals that the regularization effects make the Lagrangian small in a strip
of size~$\varepsilon^\alpha$ with~$\alpha<1$. For our purposes, it suffices to note that
as a consequence of these regularization effects, the function~$\ell$ is much smaller than
the scaling obtained from simple power counting,
\beq \label{vaccount}
\ell(x) + \s \lesssim \frac{\sigma\, \lambda^4\, m^6}{\varepsilon^2}\:.
\eeq

\subsubsection{Contributions on the Light Cone}
As explained in~\cite[Chapter~4]{cfs} and~\cite[Section~3.2]{action},
in the Minkowski vacuum there are also contributions to the Lagrangian which are
localized on the light cone. Since these contributions do not involve the mass~$m$,
power counting gives the scaling
\beq \label{vacLC}
\ell(x) + \s \simeq \sigma\, \lambda^4\: \frac{1}{\delta^8}\: \Big( \frac{\delta}{\varepsilon} \Big)^{\hat{s}}
\eeq
with an undetermined parameter~$\hat{s}$. In order to determine the possible values of this parameter,
we write the corresponding contribution to the Lagrangian 
in the formalism of the continuum limit as
\beq \label{Lxyform}
\L(x,y) \simeq \frac{\lambda^4}{\delta^8}\: (\deg =2) \: \Big( \frac{\varepsilon}{t} \Big)^{\hat{q}}\:,
\eeq
where the factor~$(\deg =2)$ is again a contribution which is localized on the light cone and
has length dimension~$l^{-4}$, i.e.\
\[ (\deg =2) =
\frac{h(\xi)}{\varepsilon t}\: \delta_\varepsilon(\xi^2)\: \epsilon(\xi^0) \:, \]
and the function~$h$ is again smooth away from the origin and bounded
from above and below, uniformly in~$\varepsilon$.
Moreover, the factor~$(\varepsilon/t)^{\hat{q}}$ with~$\hat{q} \in \N_0$
takes into account the so-called regularization expansion.
When integrating~\eqref{Lxyform} over~$y$, one must keep in mind that
the mass expansion in powers of~$1/\delta$
is admissible only if~$r \lesssim \delta^2/\varepsilon$. Therefore,
we must only integrate over the range~$0 \leq r \leq \delta^2/\varepsilon$.
We thus obtain the scaling~\eqref{vacLC} with~$\hat{s}$ given by
\beq \label{shatrange}
\hat{s} = \max \big(0, 2-2\hat{q} \big) \;\in\; \{0,2\} \:.
\eeq

\subsubsection{Contributions at the Origin}
\begin{figure}
%
\psscalebox{1.0 1.0} 
{
\begin{pspicture}(-2.5,-1.8152246)(8.33707,1.8152246)
\definecolor{colour0}{rgb}{0.8,0.8,0.8}
\pspolygon[linecolor=colour0, linewidth=0.02, fillstyle=solid,fillcolor=colour0](0.07707092,0.8047754)(0.07207093,-1.6352246)(0.24207093,-1.7302246)(0.45707092,-1.7702246)(0.7320709,-1.7802246)(1.0070709,-1.7752246)(1.2420709,-1.7452246)(1.4170709,-1.6752247)(1.457071,-1.5902246)(1.452071,0.8347754)(1.337071,0.7497754)(1.072071,0.6947754)(0.8070709,0.66977537)(0.4970709,0.6847754)(0.22707093,0.73477536)
\pscircle[linecolor=black, linewidth=0.02, fillstyle=solid,fillcolor=black, dimen=outer](0.77207094,-0.4002246){0.065}
\rput[bl](0.8770709,-0.2952246){\normalsize{$0$}}
\rput[bl](2.097071,-0.3302246){\normalsize{$\varepsilon$}}
\psellipse[linecolor=black, linewidth=0.04, dimen=outer](0.7620709,0.85977536)(0.715,0.195)
\psellipse[linecolor=black, linewidth=0.04, dimen=outer](0.76707095,-1.6202246)(0.715,0.195)
\psline[linecolor=black, linewidth=0.04](1.457071,0.8747754)(1.457071,-1.6402246)
\psline[linecolor=black, linewidth=0.04](0.067070924,0.8797754)(0.067070924,-1.6352246)
\psline[linecolor=black, linewidth=0.02](1.7570709,0.8447754)(1.8570709,0.7447754)(1.8570709,-0.1552246)(1.957071,-0.25522462)(1.8570709,-0.3552246)(1.8570709,-1.5552247)(1.7570709,-1.6552246)
\psline[linecolor=black, linewidth=0.02](0.0070709228,1.1947753)(0.10707092,1.2947754)(0.65707093,1.2947754)(0.7570709,1.3947754)(0.8570709,1.2947754)(1.3570709,1.2947754)(1.457071,1.1947753)
\rput[bl](0.5070709,1.5397754){\normalsize{$m \varepsilon^2$}}
\psline[linecolor=black, linewidth=0.04, arrowsize=0.05291667cm 2.0,arrowlength=1.4,arrowinset=0.0]{->}(3.007071,-0.0052246093)(3.007071,1.1947753)
\psline[linecolor=black, linewidth=0.04, arrowsize=0.05291667cm 2.0,arrowlength=1.4,arrowinset=0.0]{->}(3.007071,-0.0052246093)(4.207071,-0.0052246093)
\rput[bl](3.1770709,1.0397754){\normalsize{$t$}}
\rput[bl](4.0970707,0.11977539){\normalsize{$\vec{\xi}$}}
\end{pspicture}
}
\caption{Scaling of the timelike region near the origin.}
\label{figcylinder}
\end{figure}
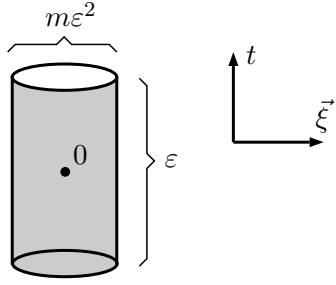

We now consider the contributions to the Lagrangian~$\L(x,y)$ at the origin (i.e.\ near the
diagonal~$x \approx y$). The scaling of these contributions was clarified only more recently
based on the connection obtained in~\cite{kilbertus} between the lowest angular momentum shell
of a Dirac system in Minkowski space and the three-dimensional Dirac sphere
in~\cite[Example~2.9]{continuum}.
In order to explain the resulting scalings, following the procedure in~\cite{reg} we consider a spherically
symmetric regularization. We set~$\xi=y-x$, write its components as~$\xi = (t, \vec{\xi})$ and 
denote~$r=|\vec{\xi}|$. Then for~$r \ll \varepsilon$, the fermionic projector can be expanded as
\[ P(t,r) \simeq\lambda \Big( \gamma^0\: \frac{1}{\varepsilon^3} + r \,\gamma^r\: \frac{1}{\varepsilon^4}
+ \frac{m}{\varepsilon^2}\: \1 \Big) \Big(1 + \O\Big( \frac{r}{\varepsilon} \Big) \Big) \:, \]
where~$\gamma^r$ denotes the radial Dirac matrices, i.e.\
\[ \gamma^r := \frac{\vec{\xi} \vec{\gamma}}{r}\:. \]
As a consequence, the closed chain scales like
\[ A = P(t,r)\, P(t,r)^* \simeq\lambda^2 \:\Big( \frac{1}{\varepsilon^6} + \frac{m}{\varepsilon^5}\: \gamma^0
+ \frac{r}{\varepsilon^7}\: i \big[ \gamma^0, \gamma^r \big]\Big) 
\Big(1 + \O\Big( \frac{r}{\varepsilon} \Big) \Big) \:. \]
The leading contribution~$\sim \lambda^2 \varepsilon^6$ is compatible with the scaling of the
eigenvalues as given in~\eqref{lamscale}. The contribution~$\sim \lambda^2 m \varepsilon^{-5}\: \gamma^0$
removes the degeneracy of the eigenvalues and gives rise to two real eigenvalues
(i.e.\ to timelike separation of~$x$ and~$y$). In particular,
\beq \label{Lxx}
\L(x,x) \simeq\lambda^4\: \frac{m^2}{\varepsilon^{10}} \:.
\eeq
The eigenvalues of the
bilinear contribution~$\sim \lambda^2 r \varepsilon^{-7}\: i \big[ \gamma^0, \gamma^r \big]$, on the other hand, 
are imaginary, giving rise to spacelike separation. At the boundary between timelike and spacelike separation
the last two contribution have the same size, i.e.
\[ \frac{m}{\varepsilon^5} \sim \frac{r}{\varepsilon^7} \]
or equivalently
\beq \label{rscale}
r \sim m \varepsilon^2 \:.
\eeq
Hence the timelike region near the origin has a cylinder-type shape, with the radius of the cylinder
being much smaller than the Planck scale. In the time direction, on the other hand, the size of this cylinder
is of the order of the regularization scale (see Figure~\ref{figcylinder}).

Using these findings, the resulting contribution to~$\ell+\s$
is computed by
\[ \ell(x) +\s \simeq\L(x,x)\: \sigma\: \varepsilon\: \big( m \varepsilon^2 \big)^3 
\overset{\eqref{Lxx}}{\sim} 
\sigma \lambda^4\: \frac{m^2}{\varepsilon^{10}} \: m^3\: \varepsilon^7 \]
and thus
\beq \label{origin}
\ell(x) +\s \simeq \frac{\sigma \,\lambda^4 \,m^5}{\varepsilon^3} \:.
\eeq

This scaling requires a few explanations. First, one should note that the resulting contribution
to~$\ell+\s$ is much {\em{smaller}} than the scaling obtained by multiplying~$\L(x,x)$ by
the volume of a four-dimensional cube of the size of the regularization length
\[ \L(x,x)\: \varepsilon^4 \simeq \lambda^4\: \frac{m^2}{\varepsilon^{6}} \:. \]
The additional scaling factor~$(\varepsilon m)^3$ in~\eqref{origin} is a result of the
effect first observed for the three-dimensional Dirac sphere
in~\cite[Example~2.9]{continuum} that bilinear contributions can be used to
``shrink'' the light cone and thus to reduce the causal action. This effect can also be understood qualitatively as the
reason why Dirac systems are favorable when minimizing the causal action.
The scaling~\eqref{rscale} also shows that visualizing regularized space-time 
as a four-dimensional lattice with lattice spacing~$\varepsilon$ is too naive.
When working with a discrete space-time, the spatial lattice must be much finer
with spacing~$m \varepsilon^2$. In the time direction, however, a lattice spacing~$\varepsilon$
seems sufficient.

We also point out that, despite this effect, the contribution near the origin~\eqref{origin}
is by a scaling factor~$(\varepsilon m)^{-1}$ {\em{larger}} than the upper bound
for the contributions away from the light cone in~\eqref{vaccount}.
This raises the question whether the contribution near the origin can be made
even smaller. There is no reason why the scaling~\eqref{origin} should be optimal.
On the other hand, at present there is no method for improving the scaling~\eqref{origin}.
Here we shall not enter the discussion of how one could possibly improve~\eqref{origin}.
Instead, we merely add the contributions~\eqref{vacLC} and~\eqref{origin}
to obtain the scaling
\beq \label{lvac}
\ell(x) +\s\simeq \sigma \,\lambda^4
\bigg( \frac{(\varepsilon m)^p}{\varepsilon^8} + \frac{1}{\delta^8}
\: \Big(\frac{\delta}{\varepsilon}\Big)^{\hat{s}} \bigg)
\eeq
with~$p \geq 5$ and~$\hat{s}$ as given by~\eqref{shatrange}.
Choosing~$p=5$ gives the scaling~\eqref{origin} obtained by regularizing Dirac sea structures.
A value of~$p>5$ can cover potential future improvements of~\eqref{origin}.

\subsubsection{Scaling of the Lagrange Multipliers}
We now compute the Lagrange Multipliers~$\kappa$ and~$\s$.
The Lagrange multiplier~$\kappa$ is obtained by minimizing the causal action,
keeping in mind the constraints. The volume constraint forces us not to change~$\sigma$.
For ease in notation, we set~$\sigma=1$. In order to build in the trace constraint, we
fix the local trace. For convenience, we arrange that the local trace is equal to one.
Thus in view of~\eqref{trx} we choose
\[ \lambda = \frac{\varepsilon^2}{m} \:. \]
Then~$\ell_\kappa$ becomes (cf.~\eqref{lvac} and~\eqref{kappacount})
\begin{align*}
&\ell_\kappa(x) +\s := \ell(x) + 
\kappa \, {\mathfrak{t}}(x) \\
&\:\simeq \sigma \,\lambda^4\: \bigg( \frac{(\varepsilon m)^p}{\varepsilon^8} + \frac{1}{\delta^8}
\: \Big(\frac{\delta}{\varepsilon}\Big)^{\hat{s}} \bigg)
+ \kappa\: \frac{\sigma\, \lambda^4}{\varepsilon^8}
= \bigg( \frac{(\varepsilon m)^p}{m^4} + \frac{1}{m^4}\: \Big( \frac{\varepsilon}{\delta} \Big)^{8-\hat{s}} \bigg) + \frac{\kappa}{m^4} \:.
\end{align*}
The remaining parameters to vary are the rest mass~$m$ and the parameter~$\delta$.
Since the parameter~$\delta$ has the purpose of generating a contribution to the mass expansion
of the neutrino sector (for details see~\cite[Chapter~4]{cfs}), the parameters~$m$ and~$\delta$
cannot be varied independently. 
Having a minimizer of the causal action implies that~$\ell_\kappa$
is minimal under such variations (for details see~\cite{lagrange}).
Moreover, the parameter~$\s$ is chosen such that this minimum is zero.
If one varies keeping the product~$m \delta$ fixed, in order for a minimum to exist
we must assume that~$p>4$ (in agreement with our finding~$p \geq 5$ in~\eqref{lvac}).
Moreover, we find
\beq
\kappa \lesssim (\varepsilon m)^p + \Big( \frac{\varepsilon}{\delta} \Big)^{8-\hat{s}} \label{kappaval} \:.
\eeq
Note that~$\kappa$ is dimensionless.
Using this value for~$\kappa$, both contributions~\eqref{kappacount}
and~\eqref{lvac} have the same scaling. Since they are both positive,
we conclude that~$\s$ scales like
\begin{align}
\s &\simeq \frac{\sigma \,\lambda^4}{\varepsilon^{8}}\:
\bigg( (\varepsilon m)^p + \Big( \frac{\varepsilon}{\delta} \Big)^{8-\hat{s}} \bigg) \:. \label{nuval}
\end{align}

\subsection{Scalings in the Presence of Matter Fields}
\subsubsection{Contributions by Matter Fields}
As worked out in detail in~\cite[Section~4.5]{cfs} and~\cite[Section~3.7]{action}, the
leading contribution of the matter fields has the form
\[ \L(x,y) \simeq \frac{1}{\delta^4} \cdot
\Big( \delta^{-2} \,R_{ij} + c\, T_{ij}[\psi] + c'\, T_{ij}[A] \Big) \:\xi^i \xi^j \:(\deg = 3) \:, \]
where~$T_{ij}$ and~$R_{ij}$ denote the energy-momentum tensor and the Ricci tensor,
respectively. These contributions cancel each other as a consequence of the Einstein equations,
but the contributions of the corresponding regularization expansion remain.
With this in mind, the relevant contributions of the matter fields to the Lagrangian are of the form
\beq \label{Lcont}
\L(x,y) \simeq \frac{\lambda^4}{\delta^4} \:T_{ij}\:\xi^i \xi^j \: (\deg=3)\: \Big( \frac{\varepsilon}{t} \Big)^{q}
= \frac{\lambda^4}{\delta^4} \:T_{ij}\:\xi^i \xi^j \: \Big( \frac{\varepsilon}{t} \Big)^{q}\: \frac{h(\xi)}{(\varepsilon t)^2}\: \delta_\varepsilon(\xi^2)\: \epsilon(\xi^0) \:,
\eeq
where we again used the notation introduced before~\eqref{znotation}.
Moreover, the factor~$(\varepsilon/t)^q$ in~\eqref{Lxyform} with~$q \geq 0$
again takes into account the contributions of the regularization expansion.
Using that the energy-momentum tensor has the length dimension~$l^{-4}$,
the contribution to the Lagrangian~\eqref{Lcont} has the desired scaling~$\sim \lambda^4 \,l^{-12}$
(see~\eqref{Lscale}). Integrating over~$y$ in the range~$0 \leq r \leq \delta^2/\varepsilon$
(see the explanation before~\eqref{vacLC}),
the resulting contribution to the function~$\ell$ has the scaling
\beq \label{lmatter}
\ell(x) \simeq \sigma \,\lambda^4 \;\frac{1}{\delta^4} \: T(x)\: \Big( \frac{\delta}{\varepsilon} \Big)^s
\eeq
with
\beq \label{spardef}
s=\max \big(0, 4-2q \big) \;\in\; \{0,2,4\}
\eeq
(for simplicity, we omit logarithms in~$\varepsilon$ and~$\delta$, because they are much less singular than
negative powers).
We also need to compute the contribution to the Lagrange multiplier term for~$\kappa$.
Matter enters the integrand of the boundedness constraint by
\beq \label{tcontri}
\sum_{i=1}^{2n} \big|\lambda_{xy}^i\big|^2 \simeq 
\lambda^4 \:T_{ij}\:\xi^i \xi^j \: (\deg=5) \:.
\eeq
Integrating over~$y$ and using~\eqref{kappaval}, we obtain the scaling
\beq \label{kappatmat}
\kappa \, {\mathfrak{t}}(x)
\lesssim \sigma \,\lambda^4 \;\bigg( (\varepsilon m)^p + \Big( \frac{\varepsilon}{\delta} \Big)^{8-\hat{s}} \bigg)
\: \frac{1}{\varepsilon^4}\: T(x) \:.
\eeq
Let us consider which of the contributions in~\eqref{lmatter} and~\eqref{kappatmat} is larger.
In the case when the second summand in~\eqref{kappatmat} dominates, i.e.\ if
\[ 
(\varepsilon m)^p \lesssim \Big( \frac{\varepsilon}{\delta} \Big)^{8-\hat{s}} \:, \]
the contribution~\eqref{lmatter} dominates because~$\varepsilon \lesssim \delta$.
In the remaining case when the first argument of the maximum dominates,
the contribution~\eqref{kappatmat}
differs from~\eqref{lmatter} by a scaling factor~$(\varepsilon m)^p\,(\delta/\varepsilon)^{4-s}$.
This scaling factor is indeed very small because, using that~$p \geq 5$,
$m \lesssim 1/\delta$ and~$s \geq 0$, we find that
\[ 
(\varepsilon m)^p \lesssim \Big( \frac{\varepsilon}{\delta} \Big)^{4-s} \:. \]
We conclude that the contribution~\eqref{kappatmat} is much {\em{smaller}} than~\eqref{lmatter} and can 
therefore be disregarded. Thus again the contribution~\eqref{lmatter} dominates.

Our findings are summarized as follows.
\begin{Prp} Under the assumption
\[ p \geq 5 \:, \]
in the presence of matter the function~$\ell_\kappa(x)$ scales like
\beq
\ell_\kappa(x) + \s \simeq \frac{\sigma \,\lambda^4}{\varepsilon^4} \:T(x) \:
\Big( \frac{\varepsilon}{\delta} \Big)^{4-s}
\label{lmat2}
\eeq
where~$\s$ again scales as in~\eqref{nuval}, and~$s$ is given by~\eqref{spardef}.
\end{Prp}
We note that the energy-momentum typically scales like
\beq \label{Tscale}
T \simeq \frac{m}{l_{\text{\tiny{macro}}}^3} \lesssim m^4 \:.
\eeq
As a consequence, the contribution on the right side of~\eqref{lmat2}
is at least by a scaling factor~$(m \delta)^4$
{\em{smaller}} than the vacuum contributions given in~\eqref{lvac}.

\begin{figure}
%
\psscalebox{1.0 1.0} 
{
\begin{pspicture}(-0.5,-1.825529)(26.475,1.825529)
\definecolor{colour0}{rgb}{0.8,0.8,0.8}
\pspolygon[linecolor=black, linewidth=0.02, fillstyle=solid,fillcolor=colour0](10.035,-1.1595916)(10.035,-1.0595915)(12.935,1.8154085)(13.12,1.8104085)(10.025,-1.2995915)
\pspolygon[linecolor=black, linewidth=0.02, fillstyle=solid,fillcolor=colour0](10.02,-1.1695915)(10.02,-1.0695915)(7.12,1.8054085)(6.935,1.8004085)(10.03,-1.3095915)
\pspolygon[linecolor=black, linewidth=0.02, fillstyle=solid,fillcolor=colour0](3.22,-1.1695915)(3.22,-1.0695915)(0.32,1.8054085)(0.135,1.8004085)(3.23,-1.3095915)
\pspolygon[linecolor=black, linewidth=0.02, fillstyle=solid,fillcolor=colour0](3.235,-1.1595916)(3.235,-1.0595915)(6.135,1.8154085)(6.32,1.8104085)(3.225,-1.2995915)
\pspolygon[linecolor=colour0, linewidth=0.02, fillstyle=solid,fillcolor=colour0](0.42,1.8104085)(6.005,1.8104085)(5.935,1.7454085)(5.86,1.6754085)(5.77,1.6004084)(5.68,1.5254085)(5.57,1.4304085)(5.415,1.3204085)(5.295,1.2254084)(5.19,1.1454085)(5.075,1.0654085)(4.89,0.9354085)(4.76,0.8554085)(4.56,0.7454085)(4.33,0.6554085)(4.14,0.5954085)(3.93,0.52040845)(3.76,0.43040848)(3.62,0.34040847)(3.55,0.27540848)(3.485,0.19040848)(3.43,0.090408474)(3.37,-0.044591524)(3.33,-0.18459152)(3.3,-0.31959152)(3.27,-0.49459153)(3.245,-0.6445915)(3.23,-0.7845915)(3.22,-1.0095916)(3.2,-0.73959154)(3.185,-0.5295915)(3.145,-0.37459153)(3.105,-0.17959152)(3.06,-0.05959152)(3.02,0.07540848)(2.93,0.23040847)(2.79,0.36540848)(2.655,0.43540847)(2.495,0.51540846)(2.315,0.58540845)(2.13,0.64540845)(1.96,0.70540845)(1.795,0.7804085)(1.45,1.0004085)(1.155,1.2104084)(0.875,1.4304085)(0.63,1.6304085)(0.515,1.7204084)
\pscircle[linecolor=black, linewidth=0.02, fillstyle=solid,fillcolor=black, dimen=outer](3.22,-1.1845915){0.065}
\psline[linecolor=black, linewidth=0.04, arrowsize=0.05291667cm 2.0,arrowlength=1.4,arrowinset=0.0]{->}(0.02,-0.18959153)(0.02,1.0104085)
\psline[linecolor=black, linewidth=0.04, arrowsize=0.05291667cm 2.0,arrowlength=1.4,arrowinset=0.0]{->}(0.02,-0.18959153)(1.22,-0.18959153)
\rput[bl](3.345,-1.7495915){\normalsize{$\L(x,y)$}}
\rput[bl](0.92,-0.0145915225){\normalsize{$\vec{\xi}$}}
\psline[linecolor=black, linewidth=0.04](3.22,-1.1895915)(0.22,1.8104085)
\psline[linecolor=black, linewidth=0.04](3.22,-1.1895915)(6.22,1.8104085)
\psline[linecolor=black, linewidth=0.04](10.02,-1.1895915)(7.02,1.8104085)
\psline[linecolor=black, linewidth=0.04](10.02,-1.1895915)(13.02,1.8104085)
\rput[bl](10.275,-1.6945915){\normalsize{$\displaystyle |xy|^2$}}
\rput[bl](2.905,-1.5345916){\normalsize{$0$}}
\pscircle[linecolor=black, linewidth=0.02, fillstyle=solid,fillcolor=black, dimen=outer](10.02,-1.1895915){0.065}
\rput[bl](9.715,-1.5245916){\normalsize{$0$}}
\rput[bl](0.17,0.7904085){\normalsize{$\xi^0$}}
\psbezier[linecolor=black, linewidth=0.02](3.22,-1.1395916)(3.2242312,-0.6088907)(3.3308089,0.10644719)(3.61,0.32040847778320314)(3.8891912,0.53436977)(4.139231,0.5661093)(4.53,0.7254085)(4.9207687,0.8847077)(5.82,1.6104084)(6.02,1.8104085)
\psbezier[linecolor=black, linewidth=0.02](3.215,-1.1395916)(3.2107687,-0.6088907)(3.104191,0.10644719)(2.825,0.32040847778320314)(2.5458088,0.53436977)(2.2957687,0.5661093)(1.905,0.7254085)(1.5142313,0.8847077)(0.615,1.6104084)(0.415,1.8104085)
\rput[bl](1.11,1.3354084){\normalsize{away from the lightcone}}
\rput[bl](0.295,-1.2395915){\normalsize{at the origin}}
\rput[bl](5.255,0.25040847){\normalsize{on the lightcone}}
\rput[bl](4.89,-0.8745915){\normalsize{$\displaystyle \frac{\lambda^4}{\delta^8} \:(\deg=2)$}}
\rput[bl](4.755,-1.7145915){\normalsize{$\displaystyle + \frac{\lambda^4}{\delta^4} \:T_{ij}\:\xi^i \xi^j \: (\deg=3)\: \Big( \frac{\varepsilon}{t} \Big)^q$}}
\rput[bl](8.215,1.4404085){\normalsize{$\displaystyle \lambda^4\, (\deg=6)$}}
\psbezier[linecolor=black, linewidth=0.02, arrowsize=0.05291667cm 2.0,arrowlength=1.4,arrowinset=0.0]{->}(2.41,-1.0845915)(2.57,-1.0595915)(2.65,-1.1495916)(2.975,-1.159591522216797)
\psbezier[linecolor=black, linewidth=0.02, arrowsize=0.05291667cm 2.0,arrowlength=1.4,arrowinset=0.0]{->}(4.785,-0.3945915)(4.625,-0.4245915)(4.52,-0.37459153)(4.34,-0.31459152221679687)
\psbezier[linecolor=black, linewidth=0.02, arrowsize=0.05291667cm 2.0,arrowlength=1.4,arrowinset=0.0]{->}(6.48,0.64540845)(6.2786956,0.8404085)(6.010652,0.7804085)(5.68,1.060408477783203)
\psbezier[linecolor=black, linewidth=0.02, arrowsize=0.05291667cm 2.0,arrowlength=1.4,arrowinset=0.0]{->}(6.859492,0.63779294)(7.05615,0.819498)(7.3151913,0.78476274)(7.6055083,0.9680240149721658)
\psbezier[linecolor=black, linewidth=0.02, arrowsize=0.05291667cm 2.0,arrowlength=1.4,arrowinset=0.0]{->}(10.085346,0.7915237)(10.315421,0.5560462)(10.834447,0.64220095)(11.149653,0.3442932339249637)
\rput[bl](8.235,0.94040847){\normalsize{$\displaystyle +\lambda^4\:T_{ij}\:\xi^i \xi^j \: (\deg=5)$}}
\end{pspicture}
}
\caption{Different contributions to the causal action in space-time.}
\label{figL}
\end{figure}
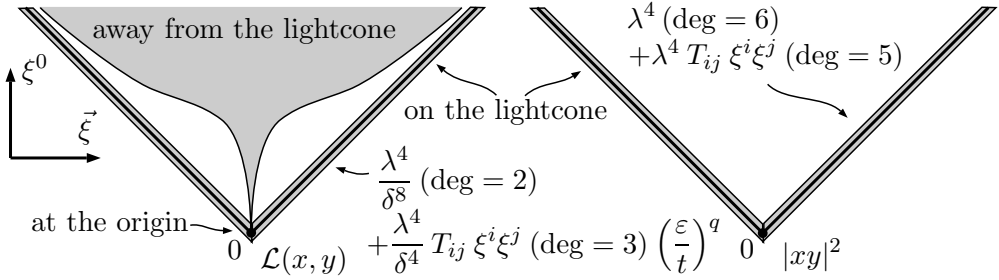

The different contributions to the causal Lagrangian and to the integrand of the
boundedness constraint are shown in Figure~\ref{figL}.
We point out that the plot of the contributions away from the light cone 
is only schematic; more details can be found in~\cite{reg}.

\subsubsection{Changing the Weight of the Measure}
Clearly, as a consequence of the matter contributions, the function~$\ell_\kappa$ in~\eqref{lmat2}
no longer satisfies the EL equations~\eqref{EL}. A simple method to arrange them is to
change the integration measure by a smooth weight function. Thus we change the measure~$\tilde{\rho}$
in~\eqref{tilrho} to
\[ \big(1 - h(x) \big) \:d\tilde{\rho}
\qquad \text{with} \qquad h(x) \simeq \varepsilon^4\;
\frac{\big(\frac{\varepsilon}{\delta} \big)^{4-s}}{
(\varepsilon m)^p + \big( \frac{\varepsilon}{\delta} \big)^{8-\hat{s}}} \: T \:. \]
Then 
\begin{align*}
\ell_\kappa(x) &= 0 \\
\ell(x) &\simeq -\frac{\sigma \,\lambda^4}{\varepsilon^8} \:
\bigg( (\varepsilon m)^p + \Big( \frac{\varepsilon}{\delta} \Big)^{8-\hat{s}} \bigg)
-\frac{\sigma \,\lambda^4}{\varepsilon^4} \:T(x) \:\Big( \frac{\varepsilon}{\delta} \Big)^{4-s} \:.
\end{align*}

\subsection{Scalings in the Killing Equation in Curved Space-Time}
We now generalize the previous results to curved space-time and use them
to determine the scalings in the Killing equation.
We first explain why in a Gaussian coordinates the results of Minkowski space
apply: When describing a curved space-time by a causal fermion system with a critical measure~$\rho$,
the constant~$c$ in the trace constraint as well as the Lagrange parameters~$\kappa$ and~$\s$
in the EL equations are global constants. These constants fix the rescaling freedom
and determine the form of the EL equations. As a consequence, in a Gaussian coordinate system
the contributions to the causal Lagrangian considered above have the same form as in Minkowski space.

In the definition of the Killing field (Definition~\ref{defkilling}), we have the situation in mind that
the vector field~$v$ describes a symmetry of the geometry, but not of the matter fields.
Therefore, when taking the directional derivatives in~\eqref{DvL}, only the contributions
of the matter fields in~\eqref{Lcont} and~\eqref{tcontri} must be taken into account
(see also Figure~\ref{figL}). Applying again~\eqref{kappaval} and using that the energy-momentum tensor
scales according to~\eqref{Tscale}, we obtain the right side of~\eqref{DvL}.

\Thanks{{{\em{Acknowledgments:}} 
E.C.\ was funded by Grant CU 338/1-1 from the Deutsche For\-schungs\-ge\-mein\-schaft.
The research of J.M.I.\ was supported by grant no.\ RTI2018-102256-B-I00 (Spain).
We would like to thank Johannes Wurm for helpful comments on the
manuscript. We are grateful to the ``Universit\"atsstiftung Hans Vielberth'' for generous support.}

\providecommand{\bysame}{\leavevmode\hbox to3em{\hrulefill}\thinspace}
\providecommand{\MR}{\relax\ifhmode\unskip\space\fi MR }
\providecommand{\MRhref}[2]{%
  \href{http://www.ams.org/mathscinet-getitem?mr=#1}{#2}
}
\providecommand{\href}[2]{#2}

\end{document}